\documentclass[runningheads]{llncs}
\usepackage{array}
\usepackage{graphicx}
\usepackage{amssymb}
\usepackage{amsmath}
\usepackage{amsfonts}
\usepackage{gensymb}
\usepackage{multirow}
\usepackage{xcolor}
\usepackage{hyperref}
\usepackage{svg}
\usepackage[percent]{overpic}

\DeclareMathOperator*{\argmin}{arg\,min}
\begin{document}
\title{HyperSpace: Hypernetworks for spacing-adaptive image segmentation}
%
%
\author{Samuel Joutard
\and
Maximilian Pietsch
\and
Raphael Prevost
}
\authorrunning{S. Joutard et al.}
%
\institute{ImFusion, Munich, Germany\\
\email{joutard@imfusion.com}}
\maketitle              
\begin{abstract}

Medical images are often acquired in different settings, requiring harmonization to adapt to the operating point of algorithms. Specifically, to standardize the physical spacing of imaging voxels in heterogeneous inference settings, images are typically resampled before being processed by deep learning models. However, down-sampling results in loss of information, whereas upsampling introduces redundant information leading to inefficient resource utilization.
To overcome these issues, we propose to condition segmentation models on the voxel spacing using hypernetworks.
Our approach allows processing images at their native resolutions or at resolutions adjusted to the hardware and time constraints at inference time.
Our experiments across multiple datasets demonstrate that our approach achieves competitive performance compared to resolution-specific models, while offering greater flexibility for the end user. This also simplifies model development, deployment and maintenance. Our code is available at \url{https://github.com/ImFusionGmbH/HyperSpace}.

\end{abstract}

\section{Introduction}

Neural network-based medical image processing has become a standard both in research and clinical settings. This approach relies on statistical learning principles, with the training set's representativeness being crucial for ensuring network robustness in clinical applications. Enhancing this representativeness can be achieved through data augmentation, which generates new training examples to broaden the training set's coverage, or through data standardization, which reduces data variability by applying a consistent pre-processing pipeline during both training and inference phases.

Unlike natural images, medical images typically come with a defined physical voxel spacing (in millimeters), essential for many processing applications. A common pre-processing practice is to use this information to standardize voxel dimensions across different images, ensuring consistency in image analysis. This resampling is crucial for ensuring that convolutional filters consistently interpret the anatomy but is a double-edged sword: reducing resolution from a higher native resolution leads to information loss, while increasing it from a lower native resolution results in the use of computing resources on partly redundant data.

\begin{figure}[t]
\centering
\includegraphics[width=\textwidth]{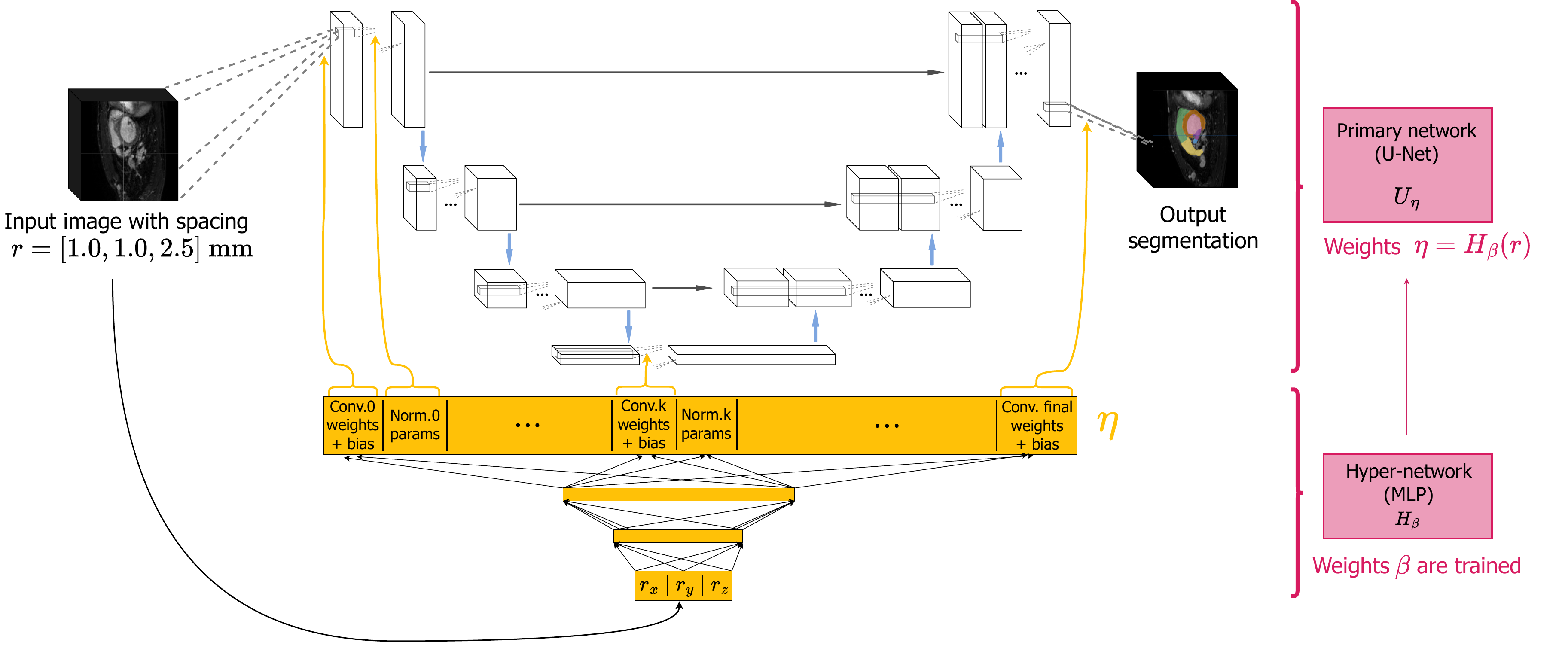}
\caption{Illustration of the proposed framework. The hyper-network $H_{\beta}$ predicts the primary network's weights $\eta$ from the image spacing. These weights (and biases) are then dispatched to their corresponding layer in the UNet that performs the segmentation.} \label{ResAgnSchema}
\end{figure}

We introduce a method to segment images at their native resolution, thus removing the need for resampling, optimizing information usage and minimizing compute requirements.
This method is based on a hypernetwork~\cite{ha2017hypernetworks}, leveraging the segmentation capabilities of U-Nets~\cite{UNet} (Section~\ref{sec.method}).
Hypernetworks are meta-architectures, where one network typically predicts all or parts of the weights of another, based on conditioning variables.
In our case, the hypernetwork takes as input the spatial spacing of the image and yields all weights of a segmentation U-Net specific to the chosen resolution.
We evaluate the performance of the method on three different datasets and segmentation tasks.
Our experiments in Section~\ref{sec.experiments} demonstrate robustness over spatial resolution and that these networks' predictions are comparable to those of models trained for and operated at a fixed image resolution. 
To provide insights into the internal structure of U-Nets generated from hypernetworks, we compare layers and models using an activation similarity measure.
We finally discuss in Section~\ref{sec.conclusion} the impact of our work from a methodological and practical perspective. 

\section{Related work}\label{sec.related_works}
 
In the more traditional image processing literature, variable resolution processing was mainly considered from the low-resolution end via the partial volume segmentation issue~\cite{1191367}. In more recent deep processing approaches,
to alleviate heterogeneous or unknown image spacings in medical image segmentation, scale equivariance has been incorporated in U-Nets, the reference architecture in medical image segmentation~\cite{Sangalli_2022_BMVC,wimmer2023scaleequivariant,yang2022scaleequivariant}. 
Such methods typically add scale as a supplementary dimension 
and apply $(D+1)$-dimensional convolutions (with D the data dimension) before reducing the scale dimension. 
Those approaches have however limitations: 
the scale dimension is discretized and
typically considered isotropic as the size of the additional dimension grows exponentially with the scale dimension. 
More fundamentally, scale-equivariance is not always desirable, as the structure size can be an essential feature (e.g. vertebrae). 

An alternative approach~\cite{rahman2023truly} processes the image in Fourier space which allows avoiding the resampling operation but not the scaling discretization, and is almost never used in medical image segmentation.
Vision transformers~\cite{dosovitskiy2021an} are popular architectures in medical image segmentation~\cite{XIAO2023104791}. 
Using world coordinates positional embeddings would make the attention mechanism spacing-adaptive.
Yet, the initial patch embedding, the integration within a U-Net architecture~\cite{chen2021transunet} 
or the use of window attention~\cite{10.1007/978-3-031-25066-8_9}
all depend on the image grid.

Another set of methods rely on statistical shape models.
Indeed, such models are typically defined with a high node density~\cite{SuinesiaputraAlbinAlbaetal.2017} or even continuously~\cite{10.1007/978-3-031-43901-8_70}. 
An agreement between the prior statistical shape model and the observed image can then be found at the native image resolution. 
Such models tend to be very robust but not extremely accurate, especially in settings of anatomical variety.\\

Our method is based on hyper-networks~\cite{ha2017hypernetworks} which have received comparatively little attention in medical image processing. 
They were first used in the context of medical image registration~\cite{melba:2022:003:hoopes} as a way to transfer the responsibility to the end-user to tune the regularization regime of the registration. 
A softer version was presented in~\cite{c-lapirn} where only the normalization layers' parameters are predicted by the hyper-network. 
Both ideas inspired the present work, but the setting of~\cite{melba:2022:003:hoopes} is more closely related as we predict the full set of resolution-specific parameters of segmentation U-Nets.\\

To summarize, our key contributions include:
\begin{enumerate}
    \item The utilization of hypernetworks for spacing-adaptive segmentation.
    \item The demonstration of this approach's effectiveness across three distinct datasets.
    \item The examination of properties of U-Nets using centered kernel alignment.
    \item Enhanced deployment flexibility of U-Net models.
\end{enumerate}

\section{Method}\label{sec.method}
\subsection{HyperSpace: Spacing-adaptive hypernetworks}

Our framework integrates two networks: a standard U-Net architecture segmentation network, $U_{\eta}$, and a hypernetwork, $H_{\beta}$, that predicts the U-nets' weights. In our study, $H_{\beta}$ is designed as a straightforward multi-layer perceptron (MLP). This MLP accepts the image voxel resolution in millimeters, $r \in \mathbb{R}^d$ (where $d$ represents the data dimension, e.g. 2 or 3), as its input and produces the weights for the U-Net, $\eta = H_\beta(r)$. 
Thus, this enables our segmentation framework to adapt to the resolution of the input image. Figure~\ref{ResAgnSchema} illustrates the overall pipeline.

During training only $\beta$ is optimized:
\begin{equation}
    \beta^* = \argmin_{\beta} \mathbb{E}_{(r, \mathcal{X}_r, \mathcal{Y}_r) \sim \mathcal{T}}\left[\mathcal{L}\left( U_{H_{\beta}(r)}(\mathcal{X}_r), \mathcal{Y}_r) \right) \right]
\end{equation}

where $\mathcal{T}$ represents the distribution of training data, $\mathcal{X}_r$ and $\mathcal{Y}_r$ are an image and its corresponding label map with voxel resolution $r$, respectively. $\mathcal{L}$ denotes a supervision loss, in our case a combination of Dice and cross-entropy loss.

In contrast to conventional settings, the expectation is defined over images, corresponding label maps, \emph{and} voxel spacings. These additional dimensions require adequate representation within the training set, further complicating the assembly of a representative dataset. Given the typically limited size of medical image datasets, we confront this challenge by employing a dedicated data augmentation strategy: each training batch is artificially resampled to a random voxel spacing, selected from the dataset's overall spacing range.

\subsection{Network analysis using network activation alignment metrics}
\label{sec.CKA}
We aim to study internal properties of U-Net instances generated by hypernetworks, going beyond performance characteristics. In particular, we investigate the similarity of activations across layers and networks using Centered Kernel Alignment (CKA)~\cite{kornblith2019similarity}, which is a similarity measure based on the Hilbert-Schmidt Independence Criterion (HSIC)~\cite{gretton2005measuring} assessing non-parametric independence between random variables. While there are theoretical concerns of CKA~\cite{davari2022reliability}, it has empirically shown consistency across varying network initializations.

HSIC evaluates similarity by comparing the squared Hilbert-Schmidt norm of the cross-covariance operator in activation spaces. CKA generalizes it by introducing invariance to isotropic scaling.
Let $X \in \mathbb{R}^{n \times p_1}$ and $Y \in \mathbb{R}^{n \times p_2}$ represent centered matrices of neural activations for the same $n$ examples but with, in general, different activation counts, $p_1$ and $p_2$, respectively. 
Linear CKA is defined as $\text{CKA}(X, Y) = \frac{\text{HSIC}(K, L)}{\sqrt{\text{HSIC}(K, K)\text{HSIC}(L, L)}}$ with $K=XX^T$, and $L=YY^T$, the activation covariance matrices. For discussion of these metrics, please see \cite{davari2022reliability}.

\section{Experiments}\label{sec.experiments}

\subsection{Datasets and Baselines}\label{sec.datasetbaselines}

Our experiments utilize public datasets of 3D MRI scans across three distinct datasets and segmentation tasks: BRATS 2021~\cite{BRATS}, SPIDER~\cite{vandergraaf2023lumbar}, and MM-WHS~\cite{ZHUANG2019101537}. We established all training and testing splits randomly. The characteristics of these datasets are summarized in Table~\ref{table.Datasets}.

\begin{table}[t]
\caption{Overview of the three public datasets used for our experiments.}\label{table.Datasets}
\centering
\begin{tabular}{|c||c|c|c|c|c|}
\hline
Dataset & Segmented Structures & Modality & Training  & Test  & Spacing Range\\
\ &   & \ & Data &  Data & (mm) \\
\hline
\hline
BRATS & Brain tumour core & T1-CE & 900 & 25 & $1.0\times1.0\times1.0$\\
SPIDER & Vertebrae, spinal cord & T1 & 447 & 10 & $[1, 4.8]\times[0.2, 1.2]^2$\\
MM-WHS & 7 Cardiac structures & T2 & 20 & 4 &  $[0.8, 1.1]^2 \times[0.9, 1.6]$\\
\hline
\end{tabular}
\end{table}

We aim at investigating whether (i)~our model demonstrates robustness across large resolution ranges, (ii)~outperforms standard data augmentation, and (iii)~provides segmentations as accurate as models trained at a fixed resolution but at a lower computational cost. 
We therefore compared our method \textbf{HyperSpace (HS)} to the following baselines:

\begin{description}
    \item[FixedSpacing (FS)]: A U-Net resampling images to a fixed resolution $r$ both at training and inference time (data harmonization). This corresponds to the standard practice. We note that recent approaches only employ data harminization during inference time~\cite{BILLOT2023102789}.
    \item[FixedSpacingNoResampling (FSNR)]: A U-Net resampling images to a fixed resolution $r$ only during training (and processing images at their native resolution at inference). This is a "dummy" baseline demonstrating the importance of considering voxel spacing.
    \item[AugmentSpacing (AS)]: A single U-Net trained with images at various voxel resolutions which is hence able to process images at native resolution at inference time (data augmentation). This would be the most straightforward  solution to deal with images at native resolution.
\end{description}

All segmentation models share the very same architecture corresponding to a 4 levels UNet with 3 convolution block per level, ReLU activations and instance normalization. The hypernetwork is a fully connected network with only 3 hidden layers, ReLU activations and a custom final activation ($:x \longrightarrow tanh(x)*5$) as a way to constraint the norm of the output weights. All models
 have been trained for 250000 iterations under identical settings with the same dataset and a combination of Dice and cross-entropy loss. 
The training procedure is very standard and all additional details are available in our code to be released.

\subsection{Performance comparisons}

\begin{table}[t]
\caption{Mean Dice score (std) on the three test sets for different resolution sub-spaces. The first interval considered corresponds to the expected resolution space. The second interval is centered around the datasets' median resolution. The third reported results consider images at their native spacing.}\label{DenseEval}
\centering
\begin{tabular}{|c||c||c|c|c|}
\hline
\multirow{2}{*}{\textbf{Datasets}} &  \multicolumn{4}{c|}{\textbf{Methods} (see Section~\ref{sec.datasetbaselines})} \\
\cline{2-5}
& FS & FSNR & AS & HyperSpace (ours)\\
\hline
\hline
BRATS $[0.5, 3.5]^3$ & 0.92 (0.06) &  0.56 (0.36) & 0.89 (0.13)  & 0.91 (0.09)\\

BRATS $[0.8, 1.2]^3$&  0.93 (0.06) &   0.88 (0.13) & 0.85 (0.15) & 0.91 (0.07)\\

\hline
 SPIDER $[1, 5]\times[0.2, 1.5]^2$& 0.89 (0.01) & 0.18 (0.03) & 0.87 (0.07) & 0.88 (0.07)\\

 SPIDER $[3, 3.5]\times[0.4, 0.8]^2$& 0.90 (0.01)  & 0.20 (0.13) & 0.88 (0.03) & 0.90 (0.02) \\

 SPIDER native& 0.91 (0.01)&  0.14 (0.01)  & 0.86 (0.08) & 0.87 (0.08)\\
\hline
 MM-WHS $[0.5, 3.5]^3$ & 0.74 (0.08)&  0.20 (0.03)  & 0.75 (0.01) & 0.79 (0.04) \\

 MM-WHS $[0.7, 1.1]^2\times[1, 1.4]$& 0.74 (0.09)  & 0.72 (0.01) & 0.77 (0.07) & 0.77 (0.07) \\

 MM-WHS native&  0.74 (0.11)  & 0.73 (0.08) & 0.76 (0.08) & 0.79 (0.06) \\
\hline
\end{tabular}
\end{table}

\begin{figure}[t!]
\centering
\includegraphics[width=0.85\textwidth]{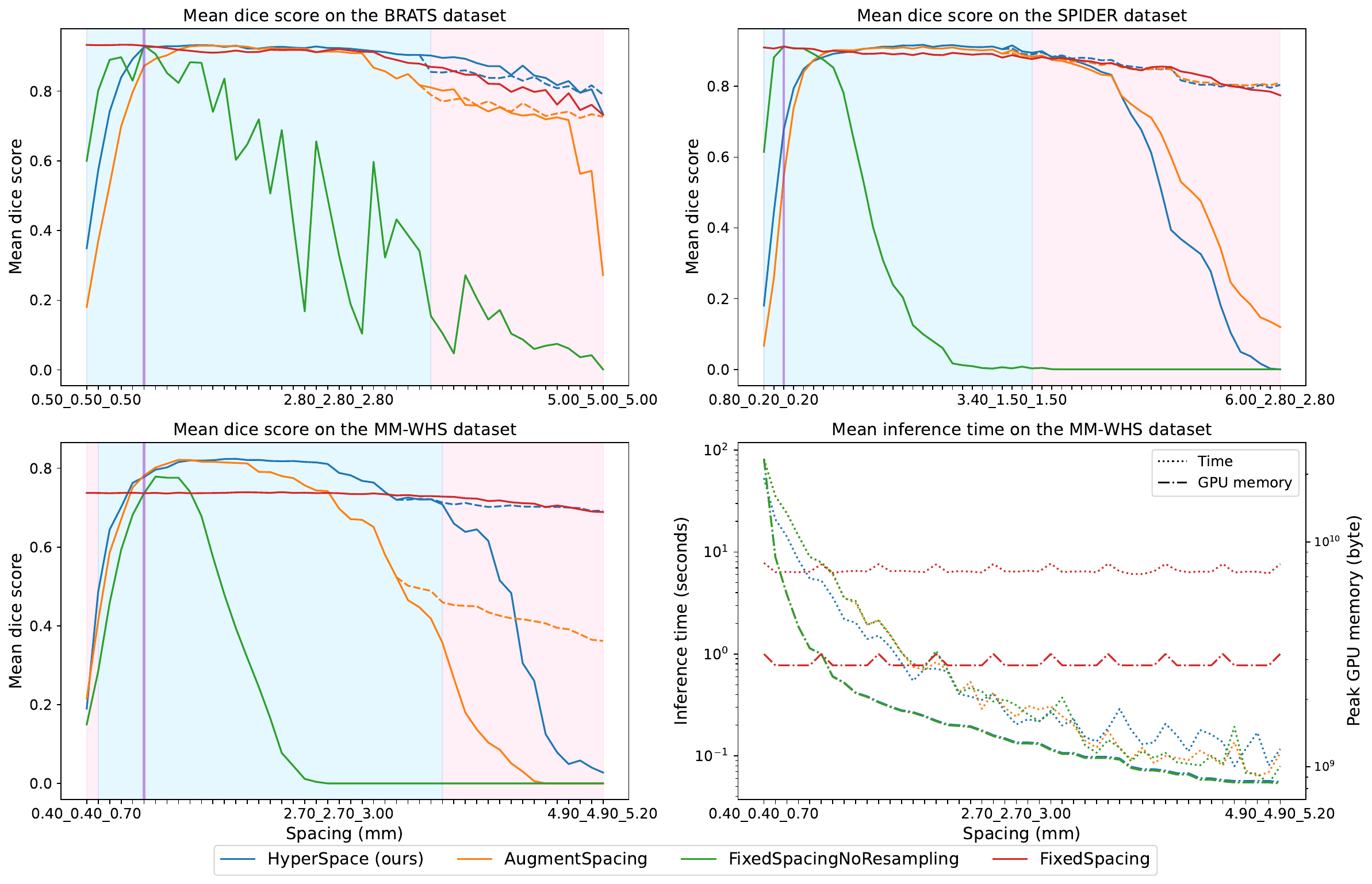}
\caption{Mean Dice score for all 3 datasets. On the bottom right, the inference runtime and peak GPU memory usage on the MM-WHS dataset are reported. The blue-shaded region correspond to expected resolution range, the pink-shaded region correspond to resolutions not seen during training. The purple vertical line indicates the resolution at which FS and FSNR were trained.} \label{DiceAcrossRes}
\end{figure}

We evaluated the different models on the held-out test sets, across various resolution intervals via Monte Carlo sampling. 
The findings are presented in Table~\ref{DenseEval}. 
Dice scores were computed at the simulated native resolution referred to as "working resolution".
As an initial sanity check, 
the poor performances of FSNR on larger resolution intervals highlights the importance of spacing considerations.
All 3 methods FS, AS and HyperSpace perform comparably on these 3 tasks. HyperSpace seems to yield better performances on the cardiac dataset where the task is presumably more complex due to the number of target structures.
However, the performance of FS comes with a significant and constant computational cost, as shown in Figure~\ref{DiceAcrossRes} (bottom right), for processing time and GPU memory usage.
Conversely, the experiment
shows that both of these requirements can be decreased by an order of magnitude by processing images at lower resolution. 
Importantly, as these measures include the full processing pipeline, including the hyper-network forward pass for HyperSpace, the additional processing cost of using a hypernetwork is negligible compared to the segmentation network's inference. 

To further investigate the behavior of the different models, performances were densely evaluated along several resolution segments, potentially going beyond the resolutions seen during training (see Figure~\ref{DiceAcrossRes}).
The hypernetwork delivers more consistent performance throughout the full expected range of resolutions, especially notable in the BRATS and MM-WHS datasets.
Indeed, we observe similar performances between AS and HyperSpace at the higher resolution end, while performances tend to diverge in favor of our method on the lower resolution side. 
We hypothesize that data augmentation is sufficient to cover smaller spacing ranges but the network weights should be adapted when the voxel size of the target structure varies significantly between images. 

However, these graphs also show that the performances of networks processing data at working resolution collapse on the highest resolution end. 
This is presumably due to the primary network's architecture being too shallow to process such high-resolution images, representing a limitation of the proposed solution as the hypernetwork currently only impacts the weights of the primary network. Being able to predict the weights, as well as modifying the architecture could allow to reliably cover larger resolution intervals. 
Furthermore, while the hypernetwork manages to extrapolates well to unseen resolutions on the BRATS dataset, the performance rapidly drops on the two others. This decrease is however not as pronounced as for AS.
These limitations could be mitigated by resampling, at test time, to the closest training resolution; as the dotted lines in Figure~\ref{DiceAcrossRes} demonstrate, this indeed improves accuracy, matching that of FS. Similarly, at very high resolution, the data can be downsampled to a resolution coherent with the segmentation network architecture.

\subsection{Internal representation analysis of generated U-Net networks}
\begin{figure}[t!]
\centering
\begin{overpic}[width=0.24\textwidth]{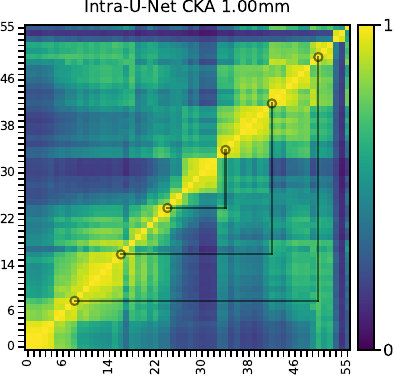}
    \put(0,95){\small{(a)}}
\end{overpic}
\begin{overpic}[width=0.24\textwidth]{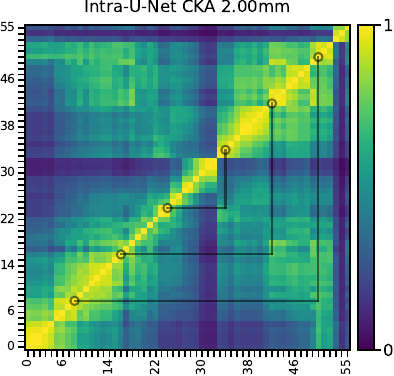}
    \put(0,95){\small{(b)}}
\end{overpic}
\begin{overpic}[width=0.248\textwidth]{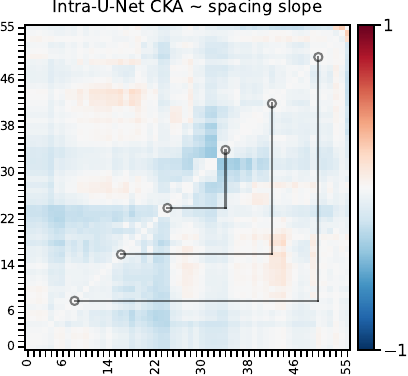}
    \put(0,95){\small{(c)}}
    \put(65,25){\color{red}\vector(1,1){1}}
    \put(30,20){\color{blue}\vector(1,1){1}}
\end{overpic}
\begin{overpic}[width=0.24\textwidth]{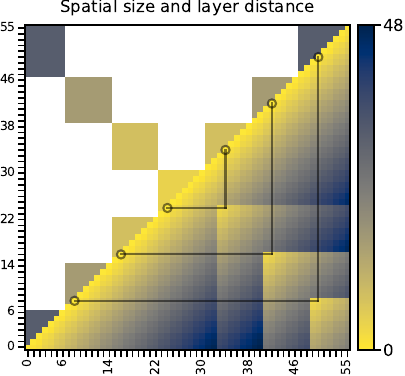}
    \put(0,95){\small{(d)}}
\end{overpic}\\[4mm]
\begin{overpic}[width=0.24\textwidth]{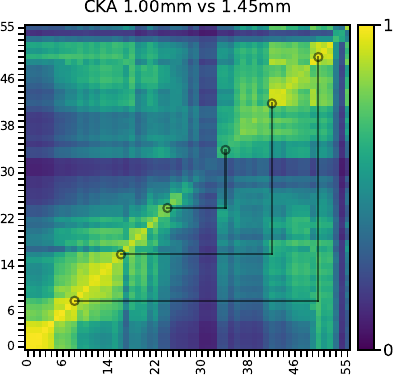}
    \put(0,96){\small{(e)}}
\end{overpic}
\begin{overpic}[width=0.24\textwidth]{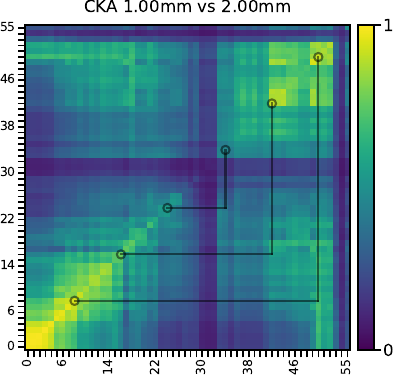}
    \put(0,96){\small{(f)}}
\end{overpic}
\begin{overpic}[width=0.248\textwidth]{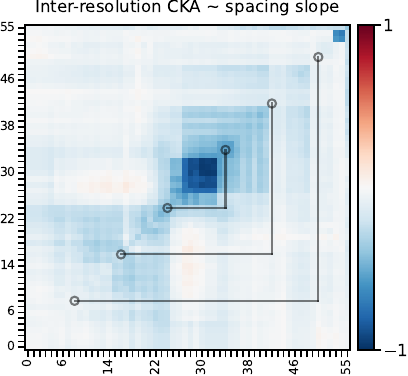}
    \put(0,95){\small{(g)}}
\end{overpic}
\begin{overpic}[width=0.24\textwidth]{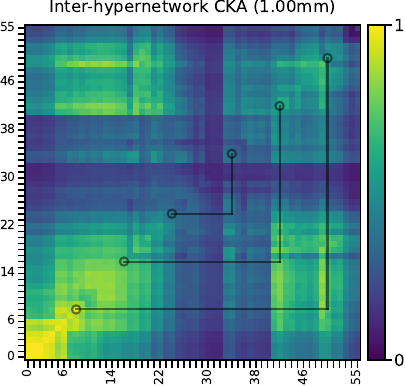}
    \put(0,96){\small{(h)}}
\end{overpic}
\caption{CKA analysis across convolution and nonlinearity layers, within resolution-specific U-Net networks (a,b) and across networks of 1mm and a coarser resolution network generated from the same hypernetwork (e,f). Plots (c,g) show the rate of change of CKA with spacing for a linear model CKA \~ spacing for the within- (c) and inter resolution-specific U-Net scores (g). For comparison, (h) shows the CKA for two U-Nets with identical spacing but from different hypernetworks. The distance between layers is shown in the lower triangle of (d) with lines indicating skip connections between U-Net branches and the upper triangle shows areas of identical spatial feature dimensions.
\label{CKAplots}}
\end{figure}

Using 437 images of the BRATS dataset for network activation generation, we compute CKA maps (see Section~\ref{sec.CKA}) for different U-Nets using activations after convolution and nonlinearity layers. 
To study the internals of hypernetworks, resolutions of images and networks are varied jointly between 0.94 and 2mm in 10 steps. 
Via resampling and cropping, we ensure that the extracted activation tensors used for CKA correspond to the same image and feature-space areas irrespective of image resolution. For compute reasons, we chose crops that correspond to a $32\times32\times32$ image volume at 1mm resolution.

We observe that intra-network CKA scores of U-Nets from the same hypernetwork are relatively stable over resolutions (Figure~\ref{CKAplots} a,b,c) and hypernetwork seed (not shown). 
In coarser resolution networks, layers between encoding and decoding branches connected through skip connections are marginally more similar (red arrow in c), while information is less well preserved across layers within the encoder and decoder branches (blue arrow in c).
While this resolution-dependent behavior is highly complex, \cite{10.5555/3495724.3496833} indicates it should converge in expectation and in the infinite width regime, to a neural network Gaussian Process with a covariance derived from the scalar product of spacings.

When comparing activations between 1mm and other resolution-specific U-Net instances from the same hypernetwork, central layers of the U-Net are the least similar across resolutions (Figure~\ref{CKAplots} g, center) indicating different information aggregation patterns in the deeper layers of the U-Net.

U-Nets from differently seeded hypernetworks share similar information in early layers (Figure~\ref{CKAplots} h) but compared to intra-network CKA and to U-Networks from the same hypernetwork (Figure~\ref{CKAplots} a), exhibit little one-to-one layer correspondence in the spatially coarse stages. 

To summarize, we can smoothly vary the resolution inputted to the hypernetwork, and the generated models, although resolution-specific, still remain closely related to each other. We hypothesize that this structure granted HyperSpace a similar convergence speed compared to other baselines despite recent evidence that HN are harder and take longer to train~\cite{ortiz2024magnitude}. Furthermore, to the best of our knowledge, we demonstrate the first CKA-based analysis of the commonly used U-Net architecture in part thanks to the well-behaved hypernetwork output space.

\section{Conclusion}\label{sec.conclusion}

In this work, we investigated the use of hyper-networks in the context of medical image segmentation as a unified model to process images at diverse spatial resolutions. 
We demonstrated on three datasets that a single hypernetwork can generate competitive U-nets for any input spacing. 
The additional cost of the proposed method is negligible and offers large computational and GPU memory benefits in low-resolution imaging settings.
We find that,
the generated U-Nets, while still exhibiting resolution-specific behaviors in the deeper layers, also show a certain consistency across resolutions in other parts. 

This work opens several further research avenues. 
We first suggest validating whether our findings that hypernetworks perform as well as fixed resolution networks, translate to other training strategies such as nnUNet~\cite{Isensee2021}. 
Then, extending the capabilities of the hyper-network for instance by using size adaptive convolution kernels~\cite{romero2022flexconv} or hyper-convolutions~\cite{ma2022hyper} in the segmentation UNet could allow to robustly cover larger resolution intervals. 
More generally, hypernetworks can be used to condition the processing of images on other image properties (e.g. contrast, patient info, etc.)

Beyond purely methodological considerations, we believe this generic and simple yet effective idea can have a practical impact. 
First, the proposed method can be leveraged when spacing information is missing or incorrect, and needs to be adjusted by the user at inference time (e.g. X-ray systems or poorly calibrated images ~\cite{Tafti2022,boese2019influence}).
Most importantly, such hypernetworks could be used as a U-Net factory, allowing the deployment of segmentation models that can be adjusted at inference time to the hardware capability or image resolution.
Their flexibility alleviates the need for training distinct models per spacing 
(see, for instance, the popular TotalSegmentator models~\cite{doi:10.1148/ryai.230024})
and would democratize access to publicly released models, irrespective of computational resources.


\bibliographystyle{splncs04}
\bibliography{bibli}

\newpage

\begin{center}
\textbf{\large [Supplementary material] HyperSpace: Hypernetworks for spacing-adaptive image segmentation}
\end{center}

\begin{figure}
\centering
\includegraphics[width=0.9\textwidth]{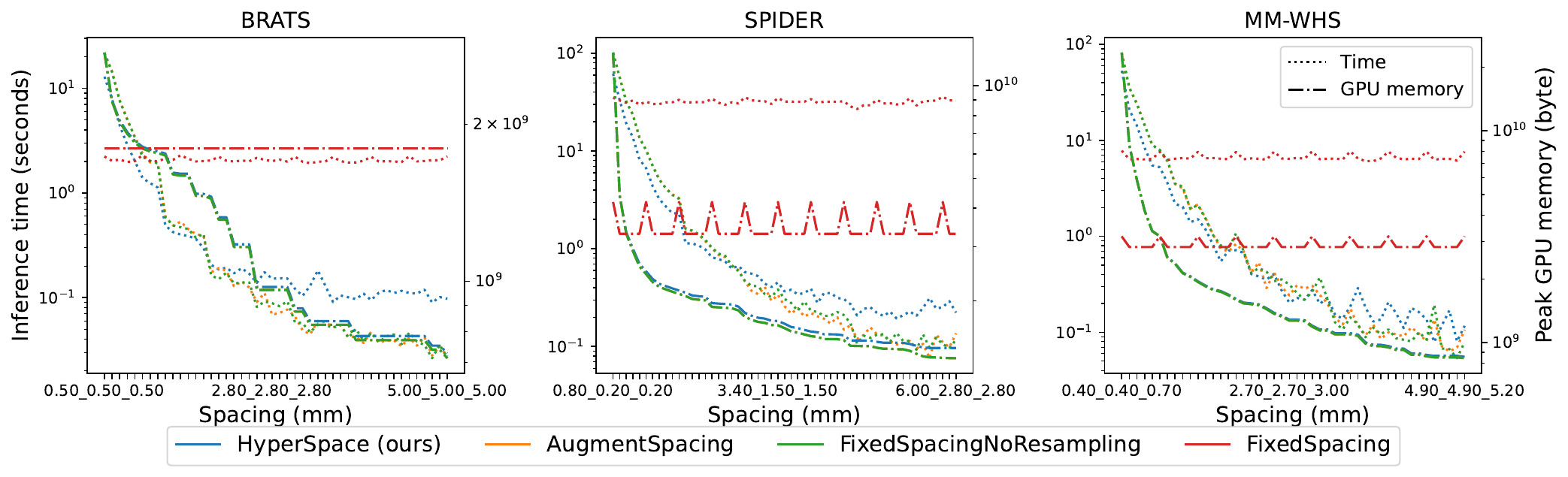}
\caption{Inference runtime and peak GPU memory usage on all 3 datasets.} \label{DiceAcrossResSeg1}
\end{figure}

\begin{figure}
\centering
\includegraphics[width=0.9\textwidth]{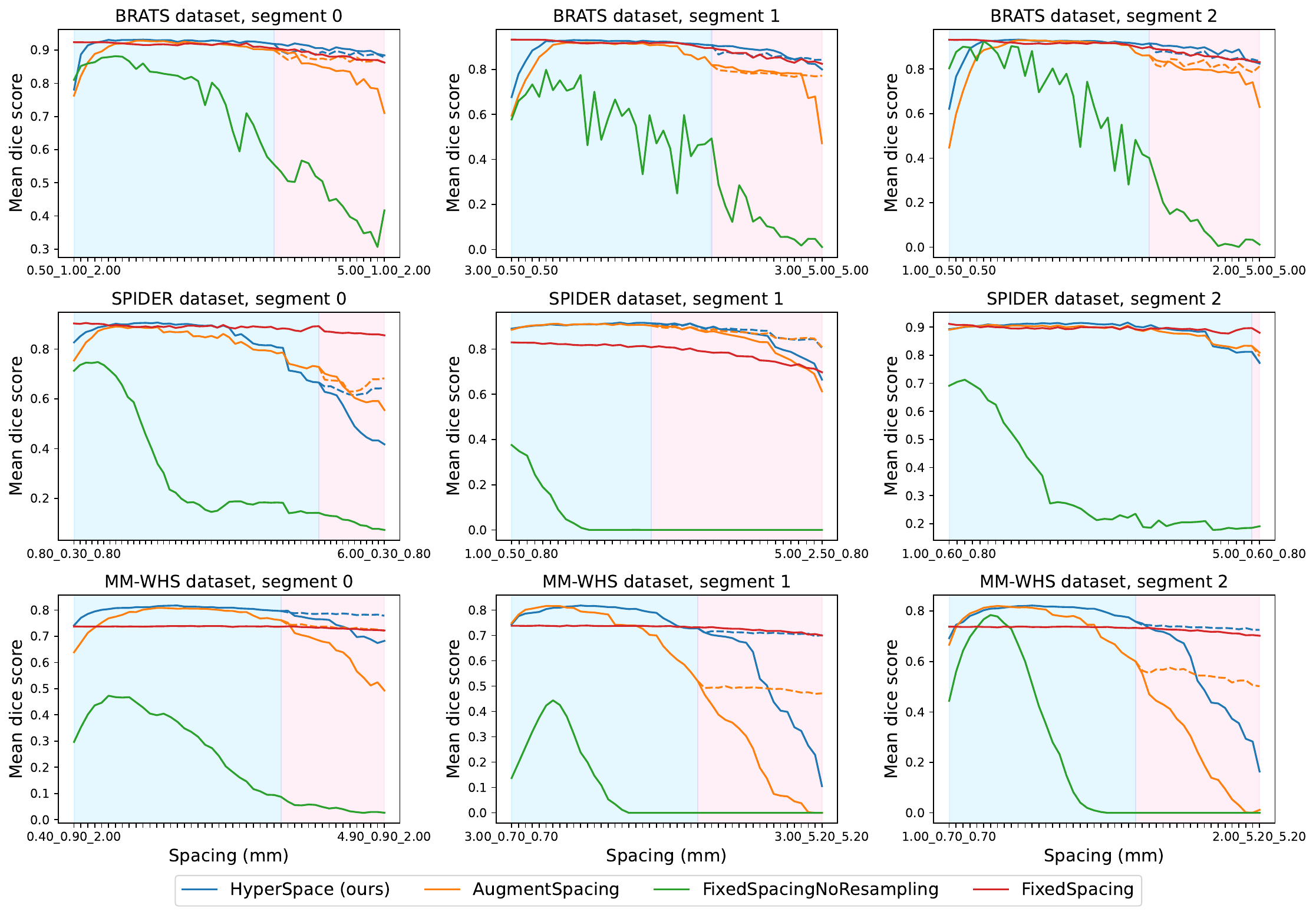}
\caption{Mean Dice score for all 3 datasets on other resolution segments.} \label{DiceAcrossOtherDirections}
\end{figure}

\begin{figure}
\includegraphics[width=\textwidth]{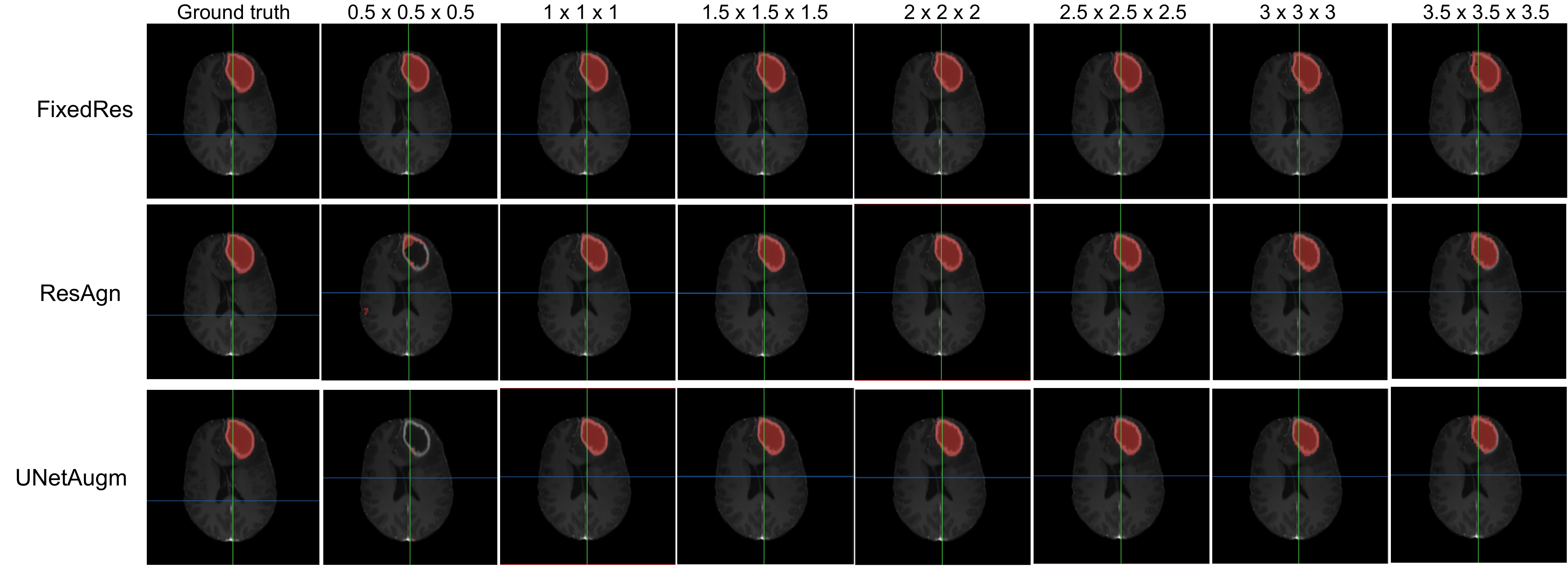}
\caption{Visualization of tumor core \fcolorbox{red}{red}{\rule{0pt}{2pt}\rule{2pt}{0pt}} segmentation on the BRATS dataset obtained from different working resolutions} \label{QualiBRATS}
\end{figure}

\begin{figure}
\includegraphics[width=\textwidth]{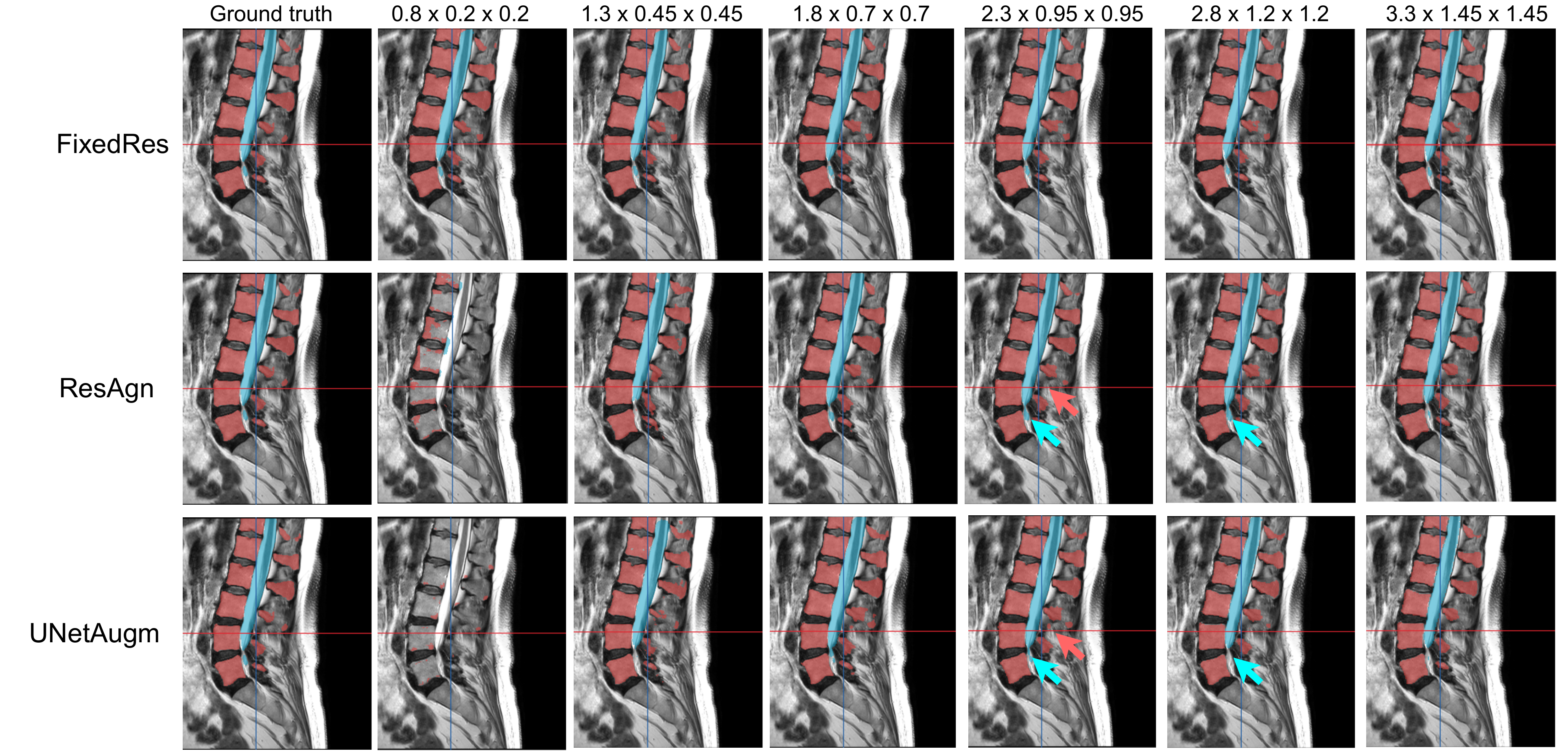}
\caption{Visualization of vertebrae \fcolorbox{red}{red}{\rule{0pt}{2pt}\rule{2pt}{0pt}} and spinal cord \fcolorbox{cyan}{cyan}{\rule{0pt}{2pt}\rule{2pt}{0pt}} segmentation on the SPIDER dataset obtained from different working resolutions} \label{Qualispine}
\end{figure}

\begin{figure}
\includegraphics[width=\textwidth]{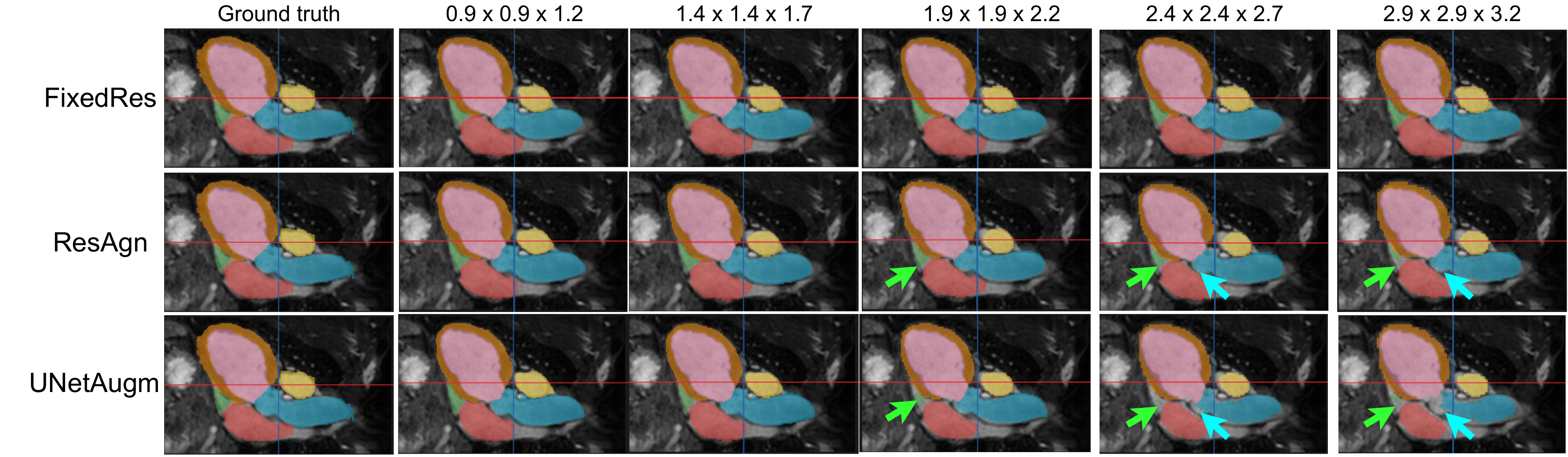}
\caption{Visualization of LV \fcolorbox{pink}{pink}{\rule{0pt}{2pt}\rule{2pt}{0pt}}, Myo \fcolorbox{orange}{orange}{\rule{0pt}{2pt}\rule{2pt}{0pt}}, AO \fcolorbox{cyan}{cyan}{\rule{0pt}{2pt}\rule{2pt}{0pt}}, RV \fcolorbox{green}{green}{\rule{0pt}{2pt}\rule{2pt}{0pt}}, PA \fcolorbox{yellow}{yellow}{\rule{0pt}{2pt}\rule{2pt}{0pt}}, LA \fcolorbox{violet}{violet}{\rule{0pt}{2pt}\rule{2pt}{0pt}}, RA \fcolorbox{red}{red}{\rule{0pt}{2pt}\rule{2pt}{0pt}} segmentation on the MM-WHS dataset obtained from different working resolutions} \label{Qualicardiac}
\end{figure}

\end{document}